\documentclass[conference]{IEEEtran}
\usepackage{fancyhdr} 
\IEEEoverridecommandlockouts
\usepackage{cite}
\usepackage{amsmath,amssymb,amsfonts}
\usepackage{algorithmic}
\usepackage{graphicx}
\usepackage{textcomp}
\usepackage{xcolor}
\usepackage{paralist}
\usepackage{enumitem}
\usepackage{threeparttable}    
\usepackage{multirow}
\usepackage{multicol}
\usepackage{array}
\usepackage{float}
\usepackage{tabularx}
\usepackage{makecell}
\usepackage{mhchem}
\usepackage{listings} 
\usepackage{graphicx} 
\usepackage{float} 
\usepackage{subfigure} 
\usepackage{caption}
\usepackage{pdfpages}
\usepackage{amsmath}
\usepackage{url} 
\usepackage[switch]{lineno}

\def\BibTeX{{\rm B\kern-.05em{\sc i\kern-.025em b}\kern-.08em
    T\kern-.1667em\lower.7ex\hbox{E}\kern-.125emX}}

\begin{document}

\title{Scaling Molecular Dynamics with \textit{ab initio} Accuracy to 149 Nanoseconds per Day\\}

\author{
\IEEEauthorblockN{
Jianxiong Li\IEEEauthorrefmark{1}\IEEEauthorrefmark{2},
Boyang Li\IEEEauthorrefmark{1}\IEEEauthorrefmark{2},
Zhuoqiang Guo\IEEEauthorrefmark{1}\IEEEauthorrefmark{2}, 
Mingzhen Li\IEEEauthorrefmark{1}\IEEEauthorrefmark{2}, 
Enji Li\IEEEauthorrefmark{1}\IEEEauthorrefmark{2}, 
Lijun Liu\IEEEauthorrefmark{3}, 
Guojun Yuan\IEEEauthorrefmark{1}\IEEEauthorrefmark{2}, \\
Zhan Wang\IEEEauthorrefmark{1}\IEEEauthorrefmark{2},  
Guangming Tan\IEEEauthorrefmark{1}\IEEEauthorrefmark{2}, and
Weile Jia\IEEEauthorrefmark{1}\IEEEauthorrefmark{2}\IEEEauthorrefmark{4} \thanks{\IEEEauthorrefmark{4}Corresponding author.}
}
\IEEEauthorblockA{\IEEEauthorrefmark{1}State Key Lab of Processors, Institute of Computing Technology, Chinese Academy of Sciences }
\IEEEauthorblockA{\IEEEauthorrefmark{2}University of Chinese Academy of Sciences }
\IEEEauthorblockA{\IEEEauthorrefmark{3}Department of Mechanical Engineering, Graduate School of Engineering, Osaka University }
\IEEEauthorblockA{\IEEEauthorrefmark{1}\{lijianxiong20g, liboyang22s, guozhuoqiang20z, limingzhen, lienji23s\}@ict.ac.cn, \{yuanguojun, wangzhan\}@ncic.ac.cn, \\ \{tgm, jiaweile\}@ict.ac.cn}
\IEEEauthorblockA{\IEEEauthorrefmark{3}liu@mech.eng.osaka-u.ac.jp}
}

\maketitle

\thispagestyle{fancy}
\lhead{}
\rhead{}
\chead{}
\lfoot{\footnotesize{
    SC24, November 17-22, 2024, Atlanta, Georgia, USA
    \newline 979-8-3503-5291-7/24/\$31.00 \copyright 2024 IEEE}
}
\rfoot{}
\cfoot{}
\renewcommand{\headrulewidth}{0pt}
\renewcommand{\footrulewidth}{0pt}

\begin{abstract}
Physical phenomena such as chemical reactions, bond breaking, and phase transition require molecular dynamics (MD) simulation with \textit{ab initio} accuracy ranging from milliseconds to microseconds. However, previous state-of-the-art neural network based MD packages such as DeePMD-kit can only reach 4.7 nanoseconds per day on the Fugaku supercomputer. In this paper, we present a novel node-based parallelization scheme to reduce communication by 81\%, then optimize the computationally intensive kernels with sve-gemm and mixed precision. Finally, we implement intra-node load balance to further improve the scalability. Numerical results on the Fugaku supercomputer show that our work has significantly improved the time-to-solution of the DeePMD-kit by a factor of 31.7x, reaching 149 nanoseconds per day on 12,000 computing nodes. This work has opened the door for millisecond simulation with ab initio accuracy within one week for the first time.

\end{abstract}

\begin{IEEEkeywords}
High Performance Computing, Molecular Dynamics, LAMMPS, Computational Science, DeePMD-kit
\end{IEEEkeywords}

\section{Introduction}

Molecular dynamics (MD) with \textit{ab initio} accuracy has been the preferred method for modeling quantum physical phenomena, including but not limited to material defects~\cite{ainsworth1984assessment}, phase transition~\cite{onuki2002phase}, and nanotechnology~\cite{raty2005growth}. However, a critical challenge in this field is the extensive physical time required to simulate these phenomena. For example, catalytic reactions typically occur within the nanosecond to microsecond range~\cite{callender2006advances}, whereas the temporal resolution of MD simulations is commonly set to approximately one femtosecond~\cite{anton}. This temporal granularity necessitates between $10^6$ to $10^9$ MD steps for such simulations.

\textit{Ab initio} molecular dynamics (AIMD) based on solving the Kohn-Sham equation has been extensively explored and studied since its inception in 1985~\cite{car1985unified}. Nonetheless, the computational cost of solving the eigenvalue problem, scaling cubically with system size ($O(N^3)$: N is the number of atoms) restricts AIMD applications to systems comprising merely thousands of atoms over picoseconds, even when leveraging heterogeneous computing architectures~\cite{jia2013analysis,JIA2013102,yanyujin2024jcst10millionatom,jia_sc11_fast}.
Neural-network-based molecular dynamics (NNMD), which was first proposed by Behler and Parrinello~\cite{behler2007generalized}, offers a new methodology to attain first-principles accuracy by employing specially designed networks to approximate the total energy and atomic forces. The computational cost of NNMD is notably reduced to linear complexity $O(N)$, representing a substantial improvement over traditional methods.
Many NNMD packages, such as SNAP~\cite{THOMPSON2015316}, SIMPLE-NN~\cite{lee2019simple}, HDNNP~\cite{Behler_2014, PhysRevLett.98.146401, behler2017first},  BIM-NN~\cite{doi:10.1021/acs.jpclett.7b01072},  ANI~\cite{smith2017ani}, CabanaMD-NNP~\cite{DESAI2022108156}, SPONGE~\cite{https://doi.org/10.1002/cjoc.202100456}, DeePMD-kit~\cite{wang2018deepmd}, Schnet~\cite{schutt2017schnet}, ACE~\cite{PhysRevB.99.014104, Lysogorskiy2021}, NequIP~\cite{Batzner2022}, DimeNet++~\cite{gasteiger_dimenet_2020, gasteiger_dimenetpp_2020} and SpookyNet~\cite{Unke2021}, Allegro~\cite{musaelian2023learning}, DPA-2~\cite{zhang2023dpa}, et al. have been developed to further refine the accuracy and computational speed of NNMD. 
We list the computational speed of typical NNMD packages in Table~\ref{tab:strong scaling}. 
For example, Allegro adapts a model decoupling and can simulate 100 million atoms on 5120 A100 GPUs~\cite{musaelian2023scaling}. LAMMPS adapts atomic cluster expansion (ACE) and reaches up to 1 nanosecond per day~\cite{nguyen2021billion}  for a 1 billion-atom carbon physical system. Among all those NNMD packages, DeePMD-kit, which preserves translational, rotational, and permutational symmetries via the embedding network, can reach up to 10 billion atoms and 4.7 nanoseconds per day on the Fugaku supercomputer~\cite{guo2022extending,jia2020pushing}. It has been widely used in the simulation of the phase transition of water~\cite{zeng2023deepmd}, absorption of \ce{N2O5}~\cite{galib2020elucidating}, and \ce{MgAlCu}  ~\cite{du2023efficient}, etc.

Despite the significant advancements NNMD has made in extending the spatial and temporal boundaries of MD simulations with \textit{ab initio} accuracy, the current packages remain insufficient in simulating complex physical phenomena such as chemical reactions in combustion and the folding of small proteins, where millisecond or even longer simulations are required. For example, the current state-of-the-art DeePMD-kit requires a minimum of 212 days to reach one-millisecond simulation on Fugaku~\cite{guo2022extending}. Thus, enhancing the time-to-solution for these NNMD packages is of paramount importance. Contrasting sharply with classical force fields like Lennard-Jones potential, NNMD packages present distinct challenges. Firstly, the scaling limit of NNMD is not yet known; specifically, whether it is feasible to allocate only a few, or perhaps even a single atom, per CPU core. Note that under such conditions, the neighboring communication between ghost regions would cross multiple layers in the 3-dimensional MPI rank topology, leading to a complex and latency-dominant communication pattern. Secondly, NNMD involves a significantly higher computational load during neural network inference, primarily due to matrix-matrix multiplications. Additionally, the NNMD packages demand additional support from neural network frameworks such as TensorFlow or PyTorch, which increase the computational latency. Moreover, the load imbalance becomes more prominent when scaling to several atoms per node. 

The current many-core supercomputers have provided hardware support even though NNMD has many challenges. Taking the Fugaku supercomputer as an example, the ARM V8 architecture provides 48 CPU cores with 3.38 TFLOPS computational power under 2.2 GHz. The 6D torus network and Tofu interconnect provide new opportunities for optimizing the scalability of NNMD codes. We also remark that other supercomputers, such as Frontier and the new Sunway, also have many-core architecture and multiple NICs per node. Hence, the same optimization strategies can be ported to other many-core systems. 

In this paper, we further extend the strong scaling limit of DeePMD-kit, a state-of-the-art NNMD code with \textit{ab initio} accuracy,  on the many-core Fugaku supercomputer. Our major contributions are as follows:
 \begin{itemize}    
    \item We propose a node-based parallelization scheme, by properly matching with the network-on-chip (NoC) ring bus and Tofu Interconnect D network (TofuD) of Fugaku to achieve high-speed intra-node data gather/scatter and inter-node peer-to-peer communication, which can reduce the overall communication overhead by 81\% percent in the strong scaling scenario.
    \item We further optimize the computationally intensive kernel with sve-gemm and mixed-precision. We also reduce the computational overhead by removing the TensorFlow framework and simplifying redundant kernel. This improves the calculation efficiency by 14.11 times.
    \item We implement an intra-node load balancing strategy by spatial decomposition on a node basis, which reduces the atomic dispersion between MPI ranks by 79.7\% and improves the overall performance by 18.5\% maximum.
    \item Numerical results show that our optimized DeePMD-kit can be $31.7$ times faster than the current state-of-the-art, reaching $149$ ns/day in simulating a copper system of $0.54$ million atoms and $68.5$ ns/day on a $0.56$ million water system on $12,000$ Fugaku nodes. As far as we know, this is the fastest MD simulation with \textit{ab initio} accuracy ever achieved. 
\end{itemize}

Although we only implement the scaling of the DeePMD-kit on Fugaku, we remark that such optimization techniques can also benefit other NNMD applications and domain-decomposition problems, as well as stencil computation. 

\begin{figure*}[t!]
  \centering
  \includegraphics[scale=0.75]{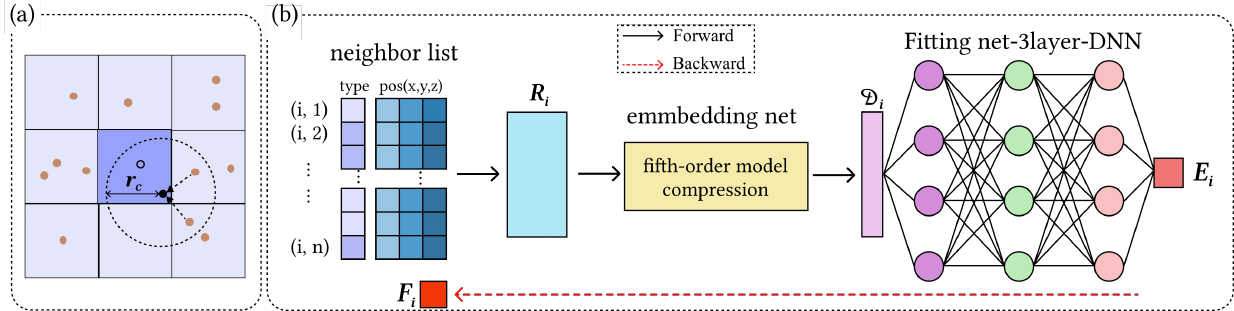}
  \caption{The sub-box and workflow of the DeePMD model.
  (a) Each MPI rank holds a sub-box, and the corresponding neighbor list within cutoff radius $r_c$ is built within the LAMMPS framework. Note that ghost regions are communicated between adjacent MPI ranks. 
  (b) The execution of the DeePMD model. Note that the atomic energy $E_i$ is carried out via forward propagation, and the atomic force $F_i$ is calculated through backward propagation. 
  }
  \label{fig:deppmd_fig}
\end{figure*}

\begin{table*}[htbp]
\caption{Performance of typical NNMD packages. The abbreviations BP and DP refer to the Behler-Parrinello scheme and Deep Potential, respectively. We remark that the DeePMD-kit is currently state-of-the-art in terms of time-to-solution.}
\label{tab:strong scaling}
\begin{threeparttable}
\begin{tabular}{llllllllll}\hline
        Work & Year & Pot & System & \#atoms & \# CPU cores & \# GPUs & Machine &  time-step & ns/day  \\ \hline
        Simple-NN~\cite{lee2019simple} & 2019 & BP & \ce{SiO2} & 14K & 80 & - & Unknown‡ &  & \\ 
        Singraber el al.~\cite{singraber2019parallel} & 2019 & BP & \ce{H2O} & 8.4K & 512 & - & VSC† & 0.5fs & 1.25  \\ 
        SNAP ML-IAP~\cite{nguyen2021billion} & 2021 & SNAP & \ce{C} & 1B & 204.6K & 27.3K & Summit & 0.5fs &1.03  \\
        Allegro~\cite{musaelian2023learning} & 2023 & Allegro  & \ce{Li3PO4} & 0.42M & - & 64 & A100 & 2fs &15.5  \\ 
        Allegro~\cite{musaelian2023learning} & 2023 & Allegro  & \ce{Ag} & 1M & - & 128 & A100 & 5fs &49.4  \\  
        DeePMD-kit~\cite{guo2022extending} (baseline) & 2022 & DP & \ce{Cu} & 13.5M & 204.6K & 27.3K & Summit & 1fs & 11.2  \\ 
        DeePMD-kit~\cite{guo2022extending} (baseline) & 2022 & DP & \ce{Cu} & 2.1M & 218.8K & - & Fugaku & 1fs & 4.7  \\ \hline
        This work & 2024 & DP & \ce{Cu} & 0.5M &  576K & - & Fugaku & 1fs & 149 \\ 
        This work & 2024 & DP & \ce{H2O} & 0.5M & 576K & - & Fugaku & 0.5fs & 68.5 \\ \hline
    \end{tabular}       
\end{threeparttable}
\end{table*}

\section{Background}

\subsection{DeePMD-kit}

Neural network-based force fields have substantially accelerated the computational speed of molecular dynamics with \textit{ab initio} accuracy. Currently, DeePMD-kit is the state-of-the-art NNMD code~\cite{jia2020pushing,guo2022extending}. DeePMD-kit mainly has two components: the LAMMPS framework and the Deep Potential (DeePMD) force field.

First, the MD operations are primarily managed by LAMMPS, which handles atom distribution, neighbor list construction, inter-process communication and atomic movement in MD simulation. Consequently,  DeePMD-kit inherits the parallelization scheme of LAMMPS, employing domain decomposition via MPI ranks. Each MPI rank obtains atoms through communication with its neighboring MPI ranks and computes forces of local atoms based on the force field. Then all MPI ranks update the atomic velocities and positions of local atoms for the next time-step and evaluate atomic physical properties, such as pressure and temperature. Additionally, DeePMD-kit adopts the load-balancing strategy of LAMMPS, ensuring an equitable distribution of computational workload among MPI ranks by adjusting sub-box borders during domain decomposition.

As shown in Fig.~\ref{fig:deppmd_fig}~(a), the atomic force and energy of atom ``i'' are determined by its local environment, i.e., its neighboring atoms. First, the local environment matrix $R_i$ within a cutoff radius $r_{c}$ is generated from the neighbor list of atom $i$. The smoothed first column of $R_i$ is then passed to the three fully connected layer network, known as the ``embedding net'', to obtain the atomic descriptor $\mathcal{D}_i$. Note that the physical symmetries, such as translational, rotational, and permutational invariance, are preserved in the atomic descriptor $\mathcal{D}_i$. Next, the atomic energy $E_i$ is evaluated via a three-layer fully connected network, named the ``fitting net'', and the total energy is carried out via summation of all atomic energy. The atomic energy $E_i$ is calculated via forward propagation, and the atomic force $F_i$ is evaluated in the backward propagation since force is the gradient of potential energy. Recently Guo et al. adopted a compression technique and have greatly reduced the computational complexity of the embedding net~\cite{guo2022extending}.
However, in the context of strong scaling on the Fugaku supercomputer, computational bottlenecks shift to the General Matrix Multiply (GEMM) kernel of the fitting net due to lower peak computing performance compared to Graphics Processing Units (GPGPU), which necessitates further optimization, particularly on many-core supercomputers.

As for the LAMMPS optimizations, X Duan, et al.~\cite{duan2020cell} introduced a cell-list-based method on the Sunway TaihuLight supercomputer to overcome the memory bottleneck for Tersoff potentials. J Li, et al.~\cite{li2023enhance} speed up the simulation performance of Lennard-Jones potential using LAMMPS, achieving a remarkable speed of 8.7 microseconds per day on the Fugaku supercomputer. However, the strong scaling of the DeePMD-kit remains challenging due to the following reasons: In the strong scaling limit where each CPU core holds a single atom, the ghost region communication must cross multiple layers, and maintaining load balance becomes critical for achieving optimal computational speed. 

\begin{figure}[t!]
  \centering
  \includegraphics[scale=0.40]{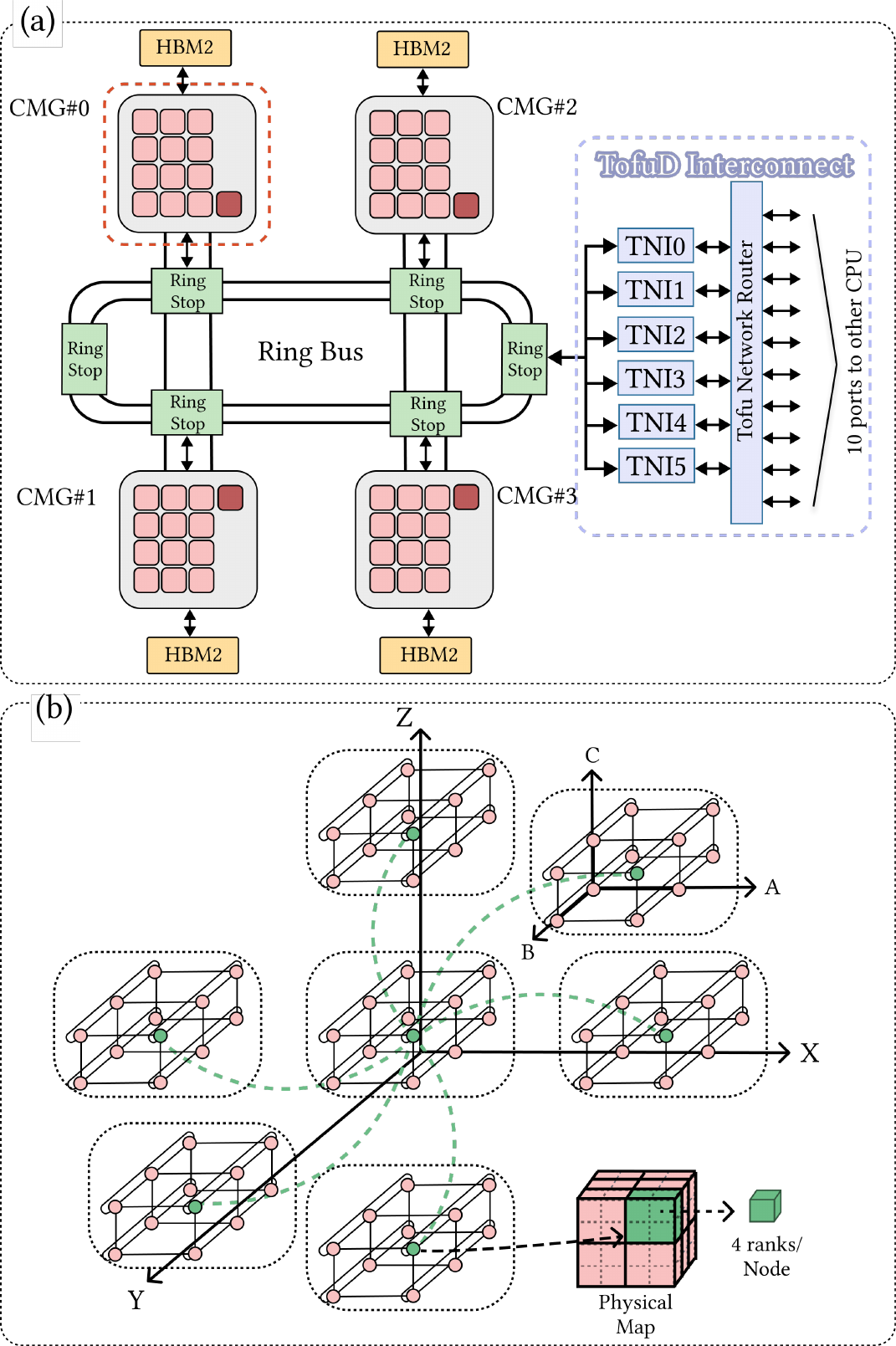}
  \caption{
  (a) The A64FX CPU has four NUMA domains and one TofuD controller, which are all connected via a high-speed network-on-chip (NoC). Each NUMA domain has 13 cores, one for OS and IO and the other 12 cores for computation. 
  Each TofuD controller is equipped with 10 ports connecting with other CPUs, as well as 6 RDMA engines (TNI) capable of simultaneously sending and receiving six packets in parallel.
  (b) 6D torus topology of the TofuD network. 12 nodes form a cell and are connected with each other through a 3D-Torus topology. Besides, each node also connects with corresponding nodes in neighboring cells. Considering that 6D torus topology can be transformed into a logical 3D torus topology, applications that use domain decomposition (such as LAMMPS) can directly map onto the system.
  }
  \label{fig:fugaku cpu}
\end{figure}

\subsection{Fugaku supercomputer}

Recently ARM based many-core supercomputers have become increasingly popular in the HPC community. 
Among them, the Fugaku supercomputer, located at the Riken Center for Computational Science, is the most powerful ARM-based supercomputer and currently ranks No.4 on the TOP500 list~\cite{top500_fugaku}. The Fugaku system comprises 158,976 nodes with a peak performance of 537 PFLOPS. Each computing node is equipped with one A64FX SoC, which consists of 52 cores with 4 dedicated for OS and IO and 48 for computation. The system is organized into a 4-NUMA structure, called the Core Memory Groups (CMGs), with each NUMA connected to its 8GB local HBM2 memory with 256GB/s bandwidth. The A64FX architecture supports 512-bit SVE operations coupled with two SIMD pipelines, enabling each core to execute 32 double-precision floating-point operations per cycle. 

One distinct feature of Fugaku is its 6D torus/mesh interconnect. As shown in Fig.~\ref{fig:fugaku cpu}~(b), each node has 10 ports, and 12 nodes form a cell. Four ports are used to connect directly with 4 nodes within the cell, and 6 ports are connected to the nodes at the same position of the six cells in [x+, x-, y+, y-, z+, z-] six directions.
The point-to-point communication latency can be as low as 0.49 microseconds. Moreover, each node has a TofuD controller facilitating remote direct memory access (RDMA) PUT/GET operation, with 6 RDMA engines per node enabling sending and receiving six packets simultaneously. Fugaku also provides the uTofu API, a low-level communication primitive, which can directly manipulate the RDMA engine to launch one-sided communication with extremely low overhead and complete asynchronous communication. The TofuD interconnect, together with the SVE extension of Fugaku, provides a significant potential for extending the strong scalability of NNMD applications such as the DeePMD-kit.

\section{Optimization}

This section introduces our main optimizations in detail, including the node-based parallelization scheme, computing optimizations, and intra-node load balance. Since each Fugaku node has four NUMA domains, we launch 4 MPI ranks on each node in the DeePMD-kit throughout this paper. Meanwhile, 12 threads are launched in each MPI rank, and each thread binds to an individual CPU core within its corresponding CMG.

\subsection{Node-based parallelization scheme}
In this section, we first analyze the parallelization schemes and then present our communication optimizations in detail.

\subsubsection{parallelization scheme analysis}
In LAMMPS, the physical system is distributed among MPI ranks via domain decomposition. Each MPI rank holds a sub-box of atoms. The local atomic forces are carried out within each MPI rank, while ghost atoms are communicated via MPI. This MPI-based parallelization scheme has been quite successful in LAMMPS, particularly in scenarios of weak scaling and strong scaling with a short cutoff radius. However, it faces several challenges when reaching the strong scaling limit of the DeePMD-kit.

\begin{itemize}    
    \item As the number of nodes increases, extensive inter-layer communication becomes a significant bottleneck due to the large cutoff of DeePMD potential and the fine-grained sub-box at the MPI rank level. This means each MPI rank has to communicate with up to 124 neighboring MPI ranks two hops away in the 3D topology, either directly or indirectly, resulting in enormous communication overhead. 
    
    \item Modern many-core supercomputers typically have multiple NUMA domains with multiple MPI ranks per node. The mapping of ghost regions for these MPI ranks on the same node often results in partial overlaps, which in turn leads to unnecessary inter-node communication.
\end{itemize}

\begin{figure}[t!]
  \centering
  \includegraphics[scale=0.4]{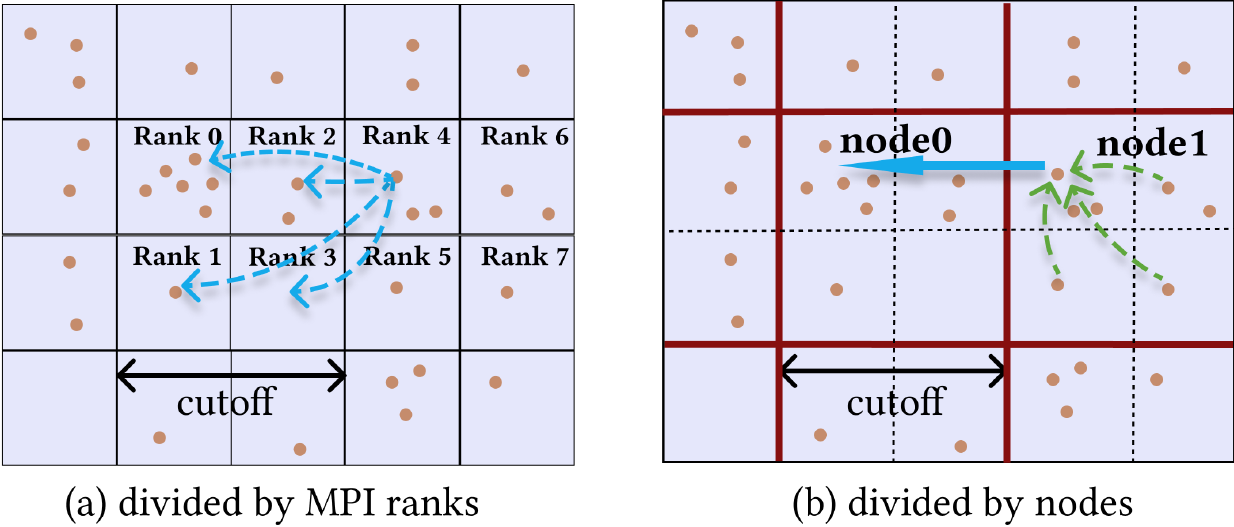}
  \caption{
  MPI-based and node-based parallelization schemes. (a) MPI-based parallelization scheme. Rank 4 and rank 5 in node 1 has to send their atoms to rank 0-3 in node 0, while rank 6 and rank 7 have to send atoms to rank 2-3. (b) Node-based parallelization scheme. The leader of node 1 gathers all atoms within the node and only has to send one message to node 0. 
  }
  \label{fig:load_imbalance}
\end{figure}

We identified an opportunity to aggregate and optimize inter-node communications by eliminating redundant transmissions, thereby reducing the overall message count. As shown in Fig. \ref{fig:load_imbalance} (a), when constructing the ghost region, rank 4 and rank 5 on node 1 have to transmit atoms to rank 0-3 on node 0, while rank 6 and rank 7 have to send atoms to rank 2 and rank 3. If we switch to the node's perspective, all atoms within node 1 are required by the MPI ranks on node 0. Thus, we can appoint a leader MPI rank, shortened for \textbf{leader}, which is responsible for gathering all the atoms of the 4 MPI ranks and sending them to the leader of neighboring nodes. Once receiving the messages, the leader MPI ranks distribute the atoms to the MPI ranks on the same node. In other words, we can communicate on the node level, termed \textbf{the node-based parallelization scheme}, which can avoid redundant data transmission. To ease the illustration, we call the 4 MPI ranks \textbf{workers}, including the leader MPI rank.

However,  the gather and scatter operations in the node-based parallelization scheme require cross-NUMA memory copy and synchronizations within the MPI ranks on the same node, resulting in significant overhead. Furthermore, inter-node communication still needs dozens of message transmissions, which requires further optimization on the Fugaku supercomputer to fully utilize the TofuD network.  We will elaborate on our optimizations in detail in the following section.

\subsubsection{communication optimization}

In the node-based parallelization scheme, the workers must copy their local atom information to the shared memory of the leader on the sending node, resulting in numerous cross-NUMA memory copies. After receiving the synchronizing signal indicating the finish of the memory copy, the leader reorganized all the data into a single packet, which is then sent to all neighboring nodes.

\begin{figure}[t!]
  \centering
  \includegraphics[scale=0.4]{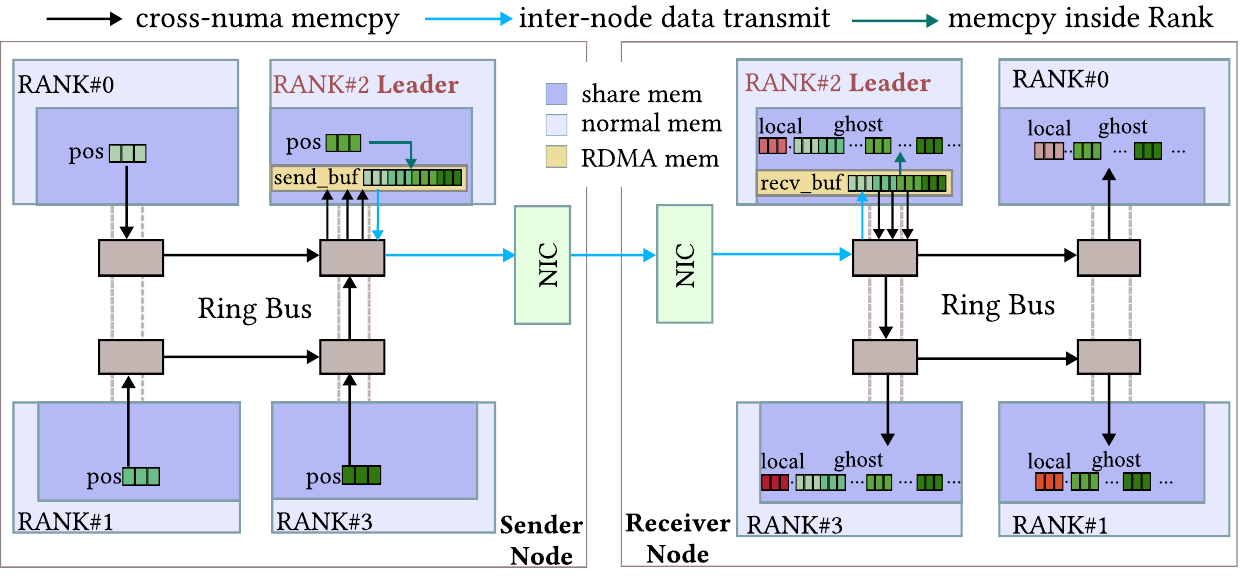}
  \caption{
  The workflow of communication in the node-based parallelization scheme. The shared memory is accessible to MPI ranks within the same node via \textit{libnuma}. Additionally, the RDMA memory is registered by \textit{libutofu} for RDMA communication and is located in the shared memory. Crucial atomic structures, including position and type information, are stored in the shared memory.
  }
  \label{fig:2layercomm}
\end{figure}

To mitigate the communication overhead, we leverage the uTofu interface to use RDMA communication, which can reduce 15\% to 27\% overhead compared to the MPI API. Since RDMA communication requires using pre-registered buffers, data must be packed to specific addresses. To reduce the memory copy of data packing, we first register the RDMA send buffer in the shared memory. Subsequently, the workers exchange their atom numbers and calculate the offset to the RDMA buffer. The workers can then directly copy their data to the corresponding position in the send buffer. Once the leader receives the completion signal of the memory copy, the RDMA communication can commence without additional data packing. The process is shown in Fig \ref{fig:2layercomm}.

Additionally, to fully utilize the 6D torus network and exploit the capabilities of the TofuD controller, peer-to-peer communication is employed to send messages directly to the leaders of the neighboring nodes. Concurrently, 6 out of 12 threads of the leader MPI rank are bound to individual TNIs to enable simultaneous communication. Note that using 6 threads for communication is our limitation due to the hardware restriction of TNI, which is not thread-safe in an MPI rank.

Upon reception, the leader unpacks the packets and scatters the split data to the shared memory of corresponding MPI ranks. Then, workers copy the received data into their internal memory. Similarly, 6 threads are employed at the receiver to monitor the message arrival signal of the corresponding TNI and do the memory copy. To further reduce the memory copy overhead, we directly put all the data structures required to be updated in the shared memory, including atom position, atom type, etc. Workers inform the leader of the offset of the atom structure in advance so that the leader can directly copy the data to the right address.

Notably, the offset of the send buffer at the sender and the offset of the atom data structure at the receiver only require to be recalculated after rebuilding the ghost region and exchanging flying atoms, which only perform once every few dozen time-steps with very little overhead.
 
By implementing the above optimizations, we can effectively reduce the communication overhead. Specifically, we require only two cross-NUMA memory copies, two intra-node synchronizations (one at the sender and another at the receiver), and a limited number of package deliveries to neighboring nodes to complete all the communication. 

Since DeePMD has to keep Newton's laws on, each MPI rank has to send back the ghost atom forces. Similar to the forward phase, the leader gathers and reduces the forces of ghost atoms before sending them to the leaders of their corresponding nodes. At the receiver, the leader further reduces all the forces of ghost atoms from all neighboring nodes and then copies the forces to the corresponding worker. Finally, all workers compute resultant forces.

Subsequently, we determine the optimal number of leaders. The A64FX uses the ring bus for intra-node communication, as well as NIC communication, with NUMA 2 and NUMA 3 situated closer to the NIC compared to NUMA 0 and NUMA 1. We bind the four MPI ranks in a node to the NUMA with the same ID. We have three options for leader selection: Rank 2 only (case 1), Rank 2 and Rank 3 as leaders (case 2), and four MPI ranks as leaders together (case 3). In case 2 and case 3, MPI ranks utilize multiple threads to copy the data to the leaders, and neighboring nodes to be communicated are equally divided among the threads of the leaders. Additionally, the leaders select the leader of the neighboring node with the same NUMA ID to send messages. 

We conducted experiments and found that communication efficiency is highest in case 3. This is because four ranks serving as leaders can simultaneously use 24 threads for data copy, force reduction, and communication. In contrast, case 1 and case 2 are limited to 6 and 12 threads, respectively. Consequently, case 3 is selected as the optimal choice for our optimized code. In fact, since all MPI ranks need to collect data from others on the same node in case 3, the data collection phase by the leader MPI ranks during communication becomes an Allgather operation.

\subsection{Computation Optimization }

In this section, we focus on our computational optimizations, including the trimming of TensorFlow, the specialized design of GEMM operators, and the mixed precision.

\subsubsection{TensorFlow Removement}

The original code adopts a hybrid parallel model in which each MPI rank initiates multiple threads, and each thread independently executes a TensorFlow session for inference. 
However, we observe a fixed overhead of approximately 4 milliseconds per session run during the DeePMD inference process. We attribute this overhead to kernel scheduling, memory management, and other operations within the TensorFlow framework. In the strong scaling scenario, where each thread handles only one or two atoms, the execution time for all computation kernels is less than 2 milliseconds, while the TensorFlow framework overhead accounts for more than 60\% of the total time. 
This substantial TensorFlow overhead significantly impacts application performance. To address this issue, we analyze the workflow of the DeePMD potential, extract all kernels participating in the force calculation from the TensorFlow framework, and rewrite the DeePMD potential accordingly. To ensure portability, we retain the TensorFlow library solely for loading model parameters.

The following optimizations are also performed during our TensorFlow removement:

First TensorFlow generates a series of kernels based on the computation graph, including gradient calculations, which can lead to redundant operations and sub-optimal performance. We streamline our code by removing unnecessary kernels.

Secondly, in the original DeePMD-kit, when simulating the multi-type system with the non-type-embedding model, the intermediate matrix is stored together. During calculations, different types of atoms in the matrix have to be extracted, with each type of atom and its neighboring atoms of different types being processed separately. This process involves multiple matrix slicing and concatenation operations, leading to excessive memory copying. 
To optimize this, we reorganize the environment matrix to pre-classify each type of atom, thereby reducing memory occupation and copying during inference.

Thirdly, the memory for all computations is allocated in the initial phase, thereby avoiding dynamic memory allocation during the simulation process.

Finally, we perform kernel fusion for all relevant kernels and cache the result of every computing step as much as possible to further enhance performance.

\subsubsection{SVE-GEMM}

The computation of 3-layer DNN takes up over 35\% of the total simulation time, which has the highest proportion of all the kernels. Previous works used the BLAS library to optimize the GEMM operation in the fitting net computation. However, in the strong scaling scenario where only one or two atoms are computed on each core, the input matrix of GEMM is tall-and-skinny, less than 3 in the M-dimension, resulting in sub-optimal performance. Therefore, we develop a high-performance GEMM based on SVE-512 for the strong scaling scenario. We make each element \textit{i} in each row of matrix A multiplies with all the elements in row \textit{i} of matrix B, and sum the result with the previous row result via MLA (vector) instruction. Consequently, GEMM can be implemented through a few vectorized operations using SVE-512 instructions (svld1, svst1, svmla\_z, etc.). After fine-tuning based on the cache size and matrix size, we observe a 1.4x performance improvement compared to the Fugaku BLAS library. 

Additionally, when computing the gradient of the fitting net, we have to calculate the product of the input gradient matrix and the transpose of the parameter matrix. Our test indicates that when the matrix size is small, the efficiency of computing the product of a matrix in non-transpose with transpose (GEMM-NT) form is halved compared to the performance of computing the product of two non-transpose (GEMM-NN) matrices. Hence, we preprocess the parameter matrix by generating its transpose in the initial phase, converting all NT calculations of the fitting net into NN calculations to improve efficiency.

\subsubsection{Mixed Precision}

The baseline work employs double-precision floating-point (double) for all calculations. However, subsequent studies have demonstrated that single-precision floating-point (fp32) suffices for accuracy requirement involving atom descriptor construction of the embedding net~\cite{lu2022dp} and the activation function of the fitting net, while the half-precision floating-point (fp16) can be utilized in the GEMM operation of the fitting net ~\cite{jia2020pushing}. Leveraging mixed-precision computation can significantly enhance performance while maintaining the accuracy. In our optimized code, we adopt the above configurations and introduce an fp16-sve-gemm implementation to further improve performance.

\subsection{Intra-node load balance}

\begin{figure}[t!]
  \centering
  \includegraphics[scale=0.40]{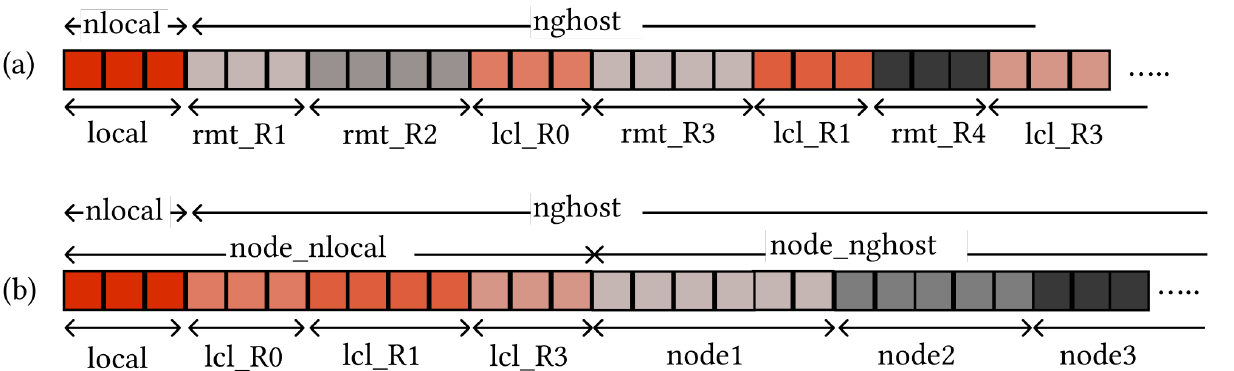}
  \caption{
  Atom organization of the two parallelization schemes in Rank 2. (a) The original parallelization scheme. $lcl\_Rx$ represents the ghost atoms from neighboring MPI rank within the same nodes, while $rmt\_Rx$ refers to the ghost atoms from neighboring MPI rank on other nodes. $nlocal$ and $nghost$ are the number of local atoms and ghost atoms, respectively. (b) The node-based parallelization scheme. $node\_nlocal$ is the local atom number within the node, while $node\_nghost$ is the ghost atom number associated with the node-box. $nodex$ represents different ghost atom groups from other nodes.
  }
  \label{fig:atom org}
\end{figure}

LAMMPS employs MPI ranks as the basic unit for domain decomposition. However, when expanding to a large number of nodes, the sub-box of each MPI rank becomes small, leading to significant differences in local atom numbers between MPI ranks, even for physical systems with uniform density, as depicted in Fig \ref{fig:load_imbalance} (a). 

The scaling limit of the DeePMD-kit is to reach 1 atom/core to achieve the fastest time-to-solution while avoiding wasting resources. However, the load imbalance across the MPI ranks might lead to a situation in which some cores have no atoms to compute, while others may have more than one atom. This can significantly slow down the computational speed. In the DeePMD-kit, the atoms are evaluated in an atom-by-atom manner, thus the evaluation of two local atoms takes nearly twice as long as that of one atom. 
Thus, the key to performance improvement is to speed up the execution time of the slowest MPI rank. Although LAMMPS offers load-balance features to adjust the sub-box border to balance the local atom count, this approach often introduces additional communication overhead and provides limited assistance for systems with uniform density.

We find that in systems with uniform density, increasing the sub-box size results in a more even distribution of atoms, as shown in Fig. \ref{fig:load_imbalance}. If we regard the simulation space of the four MPI ranks in a node as a whole, called the node-box, the differences in atom number between the node-boxes will be significantly reduced. The implementation of load balance within a node will greatly improve the efficiency of the entire system. Thus, we propose a new atomic organization strategy where the 4 MPI ranks on the same node can evenly divide the atoms in the node-box to achieve load balancing within the node.

Recall that in the previous communication optimization, all four MPI ranks are designated as leaders so that each MPI rank will gather the local atoms of the other MPI ranks on the same node before inter-node communication. Thus, all MPI ranks will get all the atoms of the node-box. After completing inter-node communication, the leaders will obtain the ghost atoms of neighboring nodes. By broadcasting all ghost atoms to all workers, every MPI rank will have identical copies of local atoms and ghost atoms of the node-box. By doing this, we can divide the node-box atom evaluation tasks evenly among four MPI ranks to realize intra-node load balance.

As shown in Fig. \ref{fig:atom org}, we change the data structure of atoms from Fig. \ref{fig:atom org}(a) to Fig. \ref{fig:atom org}(b). To maintain portability, we keep the local atoms at the very beginning, followed by the atoms from the rest of the MPI ranks on the same node. The ghost atoms from the neighboring nodes are put in order at the back. During neighboring list construction and pair force computation, 48 threads across the four MPI ranks evenly divide the node-box atoms, while in other LAMMPS phases, MPI ranks focus solely on local atoms within their sub-boxes.

It should be noted that our load-balancing implementation introduces additional memory overhead. Assuming each MPI rank has a sub-box side length of $a$ and a cutoff length of $r$, and for simplicity, let the atom density be $1$ which makes the atom number equal to the volume of the box. 
Thus, the local atom number is $a^3$. 
For the original parallelization scheme, the number of ghost atoms inside each MPI rank is shown in Equation \ref{con:origin_nghost} while in our load-balance method, the total ghost atom number is represented in Equation \ref{con:lb_nghost}.
\begin{equation}
nghost_{bs} = (a+2r)^3-a^3
\label{con:origin_nghost}
\end{equation}
\begin{equation}
nghost_{lb} = (2a+2r)*(2a+2r)*(a+2r)-a^3
\label{con:lb_nghost}
\end{equation}

In the strong scaling scenario, the sub-box side length can be $0.5rcut$ for potentials with a large cutoff radius. So, considering the case where $a=0.5r$, the number of $nghost$ in our load-balance approach is approximately $1.44$ times that of the original one. However, given that the typical number of atoms per node is usually below 100 in the strong scaling scenario, the additional atoms we introduce only add a few dozen kilobytes of memory occupation and memory copy, resulting in very little impact on the overall efficiency.

\subsection{Other Optimization }

\subsubsection{RDMA memory pool}
RDMA network interface cards require storing the connection information and the registered memory region in their hardware cache for recording connections and performing address translation. However, due to the limited cache capacity on the NIC,  the overflow of registered addresses or connections will result in excess data being replaced in the main memory. Consequently, when a message arrives, and the connection information or registered address is not found in the NIC cache, it must be read from the main memory, significantly increasing communication overhead. In the inter-node communication, we use the p2p communication for high efficiency, which results in a large number of connections. To mitigate the issue, we employ a memory pool to manage all buffers and register the large memory for RDMA communication only, effectively circumventing the NIC scaling problem.

\subsubsection{Threadpool}

The original LAMMPS and DeePMD use OpenMP for parallel computation. However, due to the significant overhead between parallel blocks introduced by the thread management of OpenMP, we employ the threadpool for parallel computation and thread management, just as the recent work of J Li, et al. ~\cite{li2023enhance}, which can make threads always running to further enhance message rate and computation efficiency. 

\section{EVALUATION}

Two typical physical systems, copper atoms and water molecules, are used as benchmarks to evaluate the performance of our optimized code. Water systems, in particular, present a formidable challenge in molecule dynamics simulation due to the intricate balance between weak non-covalent intermolecular interactions, thermal effects, and nuclear quantum effects~\cite{ko2019_water,distasio2014_water,chen2017_water}. While copper, as a representative simple metal, remains challenging for predicting properties such as surface formation energy and stacking fault energies with traditional empirical force fields (EFFs).
  
We set the cutoff radius for the water and copper systems to 6 \AA \ and 8 \AA, respectively. The neighboring atom numbers of hydrogen, oxygen, and copper atoms are 46, 92, and 512 respectively. The size of the fitting net is (240, 240, 240). The time-step is set to be 0.5 fs for water and 1.0 fs for copper. The skin for the neighboring lists is 2 \AA, which has to be rebuilt every 50 time-steps. 

In terms of the environment configuration, we launch 4 MPI ranks per node with the CPU frequency of 2.2GHz. The software language environment is tcsds-1.2.38, encompassing the compiler and the BLAS library. Due to Fugaku BLAS lacking pthread support, we utilize OpenBLAS-0.3.26 in our threadpool implementation, and Fugaku BLAS in the OpenMP version. The SVE-GEMM optimization is activated when the M dimension of the input matrix is less than or equal to 3. In other cases, the BLAS library is used.  

Our optimized code is based on the LAMMPS version of December 22, 2022, and the DeePMD-kit version of 2.0.3-Fugaku. The TensorFlow version is 2.2.0, consistent with the baseline code. The baseline results presented below are from the base code. 

\subsection{Accuracy}

We begin by evaluating accuracy. Table. \ref{tab:error_energy} presents the energy and force error of a single step compared with AIMD results with different precision. "Double" indicates calculations performed entirely under double precision. "MIX-fp32" adjusts embedding net and fitting net calculations to single precision, while other operations remain double precision. "MIX-fp16" further converts the GEMM operations in the first layer of the fitting net to half precision. The results demonstrate that mixed precision maintains AIMD-level accuracy.

\begin{figure}[tb!]
  \centering
  \includegraphics[scale=0.68]{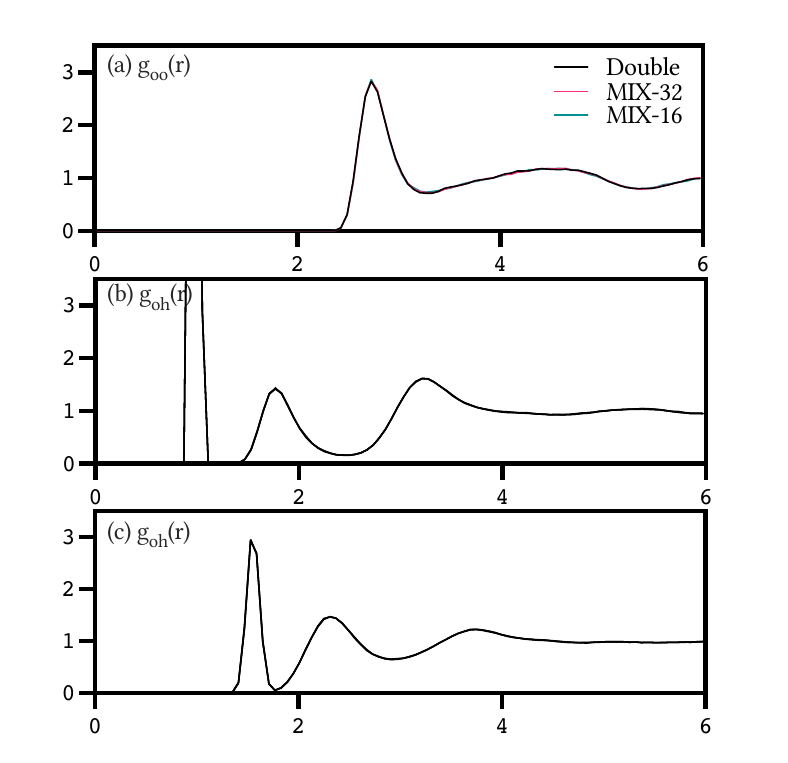}
  \caption{
  The radial distribution functions of the water system under double, MIX-fp32, and MIX-fp16 precision. The three curves overlap, proving our accuracy aligns with AIMD.
  }
  \label{fig:rdf}
\end{figure}

Fig. \ref{fig:rdf} illustrates the radial distribution functions (RDF) of the water system, which is used to characterize the water structures ~\cite{zhang2018deep}. The three curves overlap perfectly, which further proves that our accuracy keeps the  AIMD accuracy.

\begin{table}[t]
  \begin{center}
    \caption{Error of the energy and force for one time-step}
    \label{tab:error_energy}
    \begin{threeparttable}
    \begin{tabular}{c c c}\hline
      
      \textbf{Precision}     & \textbf{Error in energy [eV/atom]}  & \textbf{Error in force [eV/A]} \\
      \hline
      Double                & $1.6\times10^{-3} $              & $4.4\times10^{-2} $  \\
      MIX-fp32                 & $1.6\times10^{-3} $      & $4.4\times10^{-2} $  \\
      MIX-fp16            & $4.0\times10^{-3} $             & $4.4\times10^{-2} $ \\ \hline
    \end{tabular}
    \begin{tablenotes}    
        \footnotesize            
        \item[1] The error is compared with AIMD result.   
    \end{tablenotes}         
  \end{threeparttable}
  \end{center}
\end{table}

\subsection{Step-by-step conmmunication optimization}

\begin{figure}[t]
  \centering
  \includegraphics[scale=0.9]{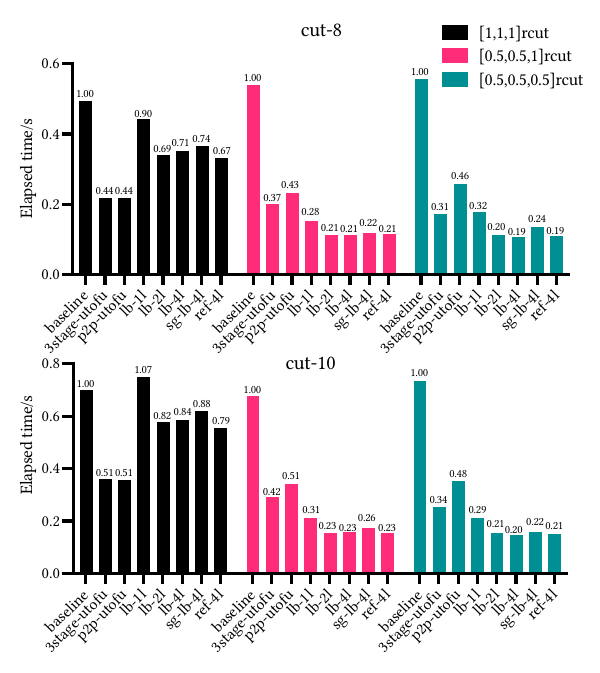}
  \caption{
  Step-by-step communication results on 96 nodes. The baseline is the original MPI-based 3-stage pattern in LAMMPS. Other tests leverage the uTofu interface. \textit{3stage-utofu} and \textit{p2p-utofu} are the 3-stage pattern and p2p pattern with the uTofu interface. \textit{lb-1l}, \textit{lb-2l}, \textit{lb-4l} represent the node-based parallelization scheme of the load balance version with 1, 2, and 4 leaders, respectively. \textit{sg-lb-4l} is the node-based parallelization scheme with 4 leader MPI ranks where each leader only communicates with a single thread. \textit{ref-4l} is the node-based parallelization scheme that preserves the original atomic organization. 
  }     
  \label{fig:coommunication_opt}
\end{figure}

In the communication evaluation, we compare the performance of the node-based parallelization scheme and two widely used communication patterns, the 3-stage pattern and the p2p pattern~\cite{li2023enhance,3-stage_pattern_desmond}, on 96 nodes with a topology of 4$\times$6$\times$4. The cutoff radiuses are 8 \AA\ and 10 \AA.
The $[1,1,1]rcut$, $[0,5,0,5,1]rcut$, and $[0,5,0,5,0,5]rcut$ denote the sub-box side length of each MPI rank.

The 3-stage pattern is LAMMPS' basic communication pattern, involving communication with neighboring MPI ranks in three opposite directions for $N$ times successively, where $N$ is the communication layer of the specific direction. In the three cases mentioned, the 3-stage pattern requires successively communicating 3, 5, and 6 times of all directions, respectively, with 2 messages each time.

The p2p pattern is another widely used communication pattern, which makes each MPI rank communicate directly with all neighboring MPI ranks. For the above 3 cases, each MPI rank has to communicate with 26, 74, and 124 neighboring MPI ranks, respectively.

As for the node-based parallelization scheme, we utilize the load-balanced communication version, which needs the leader to copy all the ghost atoms to workers. For the three cases, each node has to communicate with 26, 26, and 44 neighboring nodes, respectively, and the average messages to be transmitted in one MPI rank are 6, 6, and 11, respectively.

As shown in Fig.~\ref{fig:coommunication_opt} \textit{3stage-utofu}, \textit{p2p-utofu} and \textit{lb-4l} bars, the 3-stage and the p2p communication patterns outperform the node-based parallelization scheme when the box size is $[1,1,1]rcut$. This is because the node-based parallelization scheme requires additional synchronization and memory copy. In this case, latency is not the bottleneck; instead, bandwidth becomes the bottleneck due to the large message size in this case. Remarkably, our node-based parallelization scheme outperforms the other two communication patterns when communication spans two layers in the remaining two sub-box size configurations. This demonstrates that our communication method is more efficient in the strong scaling scenario.

Then, we test the different numbers of leader nodes and the results are shown in Fig. \ref{fig:coommunication_opt} \textit{lb-1l}, \textit{lb-2l} and \textit{lb-4l}. When all four MPI ranks are set to be leaders, network performance and NoC bandwidth are maximally utilized since more communication threads handle both communication and memory copy tasks concurrently. Meanwhile, we also test the case without multi-threading communication, shown in Fig. \ref{fig:coommunication_opt} \textit{sg-lb-4l} bars. A 10\% to 26\% performance reduction compared to the multi-threading version, which demonstrates that equipping four threads (each thread per MPI rank) can not fully utilize NIC and NoC performance.

Next, we examine the impact of additional memory copies in the load balance version on communication efficiency. As shown in Fig. \ref{fig:coommunication_opt} \textit{lb-4l} and \textit{ref-4l} bars, we observe that the additional memory copy doesn't affect the communication efficiency as expected, because it is a few kilobytes in total and is negligible compared to the NoC bandwidth.

\begin{figure}[t]
  \centering
  \includegraphics[width=0.75\linewidth]{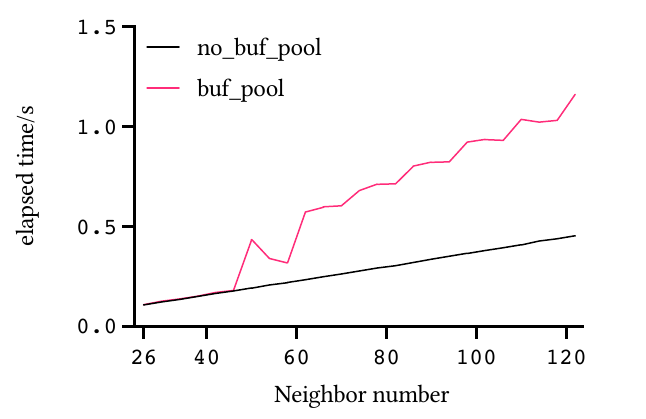}
  \caption{
  The communication time with different numbers of neighbors over 10k iterations, with the payload size of 8 bytes. The black line stands for using the memory pool. The red line represents registering the RDMA address for each neighbor severally.
  }
  \label{fig:nic_scaling}
\end{figure}

Finally, we conduct tests on our RDMA memory pool method. In the non-memory-pool version, two buffers are registered for each neighbor communication: one for receiving and one for sending. The memory-pool version registers a single large block of memory, which is sufficient for all neighbor communications, and each communication utilizes a portion of this memory block through offsets. Additionally, each MPI rank uses 6 TNIs for communication, and the messages to neighbors are sent in turn on these TNIs. As shown in Fig. \ref{fig:nic_scaling}, the communication time has linear growth in the memory-pool version, while it begins to decrease at the neighbor number of 44 in the non-memory-pool version. The irregular curve is attributed to cache scheduling strategies. Notably, our memory pool maintains excellent performance even when the neighbor number scales up to 124.

\begin{figure}[t]
  \centering
  \includegraphics[width=\linewidth]{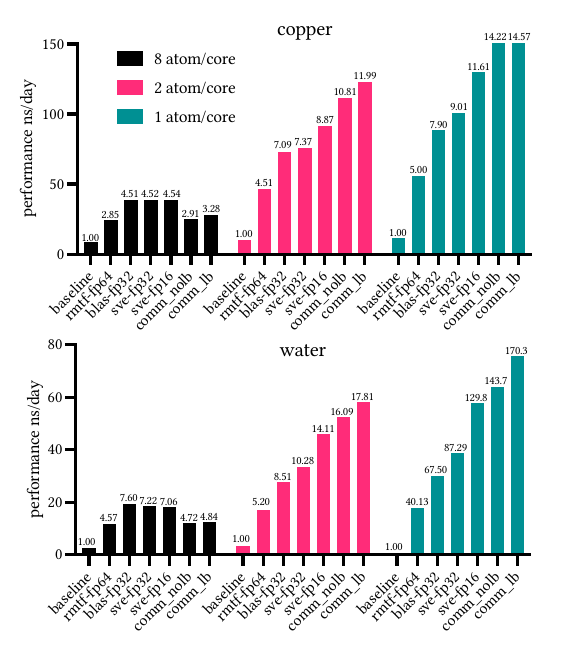}
  \caption{
  Step-by-step computation optimization results on 96 nodes over 100 time-steps. 1, 2, and 8 atoms/core represent 12, 24, and 96 atoms/MPI rank. \textit{rmtf-fp64} corresponds to the result after removing the TensorFlow framework and simplifying kernels. \textit{blas-fp32} is the result of using MIX-fp32 precision. \textit{sve-fp32} is the result of using sve-gemm approach. \textit{sve-fp16} is the result of using sve-gemm with fp16 precision. \textit{comm\_nolb} refers to the result with the node-based parallelization scheme and threadpool while not enabling intra-node load balance. \textit{comm\_lb} is the result of enabling intra-node load balance further.
  }
  \label{fig:compute_opt}
\end{figure}

\subsection{Step-by-step computing optimization}

We conduct the experiments with the following three configurations: 1 atom/core, 2 atoms/core, and 8 atoms/core, which is 0.93, 1.875, and 7.5 atoms/core on average for the copper system, and 0.968, 1.94, and 7.75 atoms/core for the water system in practice. The simulations run for 100 time-steps, plus a 10-step warmup, on 96 nodes with a topology of 4$\times$6$\times$4. The results are shown in Fig. \ref{fig:compute_opt}. The first five experiments in each configuration use OpenMP version with Fugaku BLAS library, while the last two experiments use threadpool version with OpenBLAS.

In the baseline code, we utilize the configuration of 16 MPI ranks per node and 3 threads per MPI rank, which yields the highest performance. In our optimized code, we retain utilizing 4 MPI ranks per node and 12 threads per MPI rank.

The 1-2 atoms/core configuration represents the strong scaling scenario. We also bring in the 8 atoms/core configuration to test the performance in the common scenario. Note that we do not implement specialized optimizations for the 8 atoms/core scenario as it falls beyond the scope of this paper. Consequently, the performance of sve-gemm optimizations for the 8 atoms/core setting shows no improvement. Since the \textit{comm\_nolb} and \textit{comm\_lb} version use OpenBLAS without optimization for the A64FX CPU and low-efficiency communication algorithm, the performance even decreased. Besides, we observe poor performance in the water system when the atom number is 1 atom/core in the baseline code. To ensure rigor, we do not include the enhancement ratio derived from the water system of 1 atom/core in our paper.

As shown in Fig. \ref{fig:compute_opt} \textit{rmtf-fp64} bars, we observe a significant 5.20x performance improvement at maximum by removing the TensorFlow framework, streamlining the kernel, and converting GEMM-NT to GEMM-NN. It shows that in the case of strong scaling, there is little optimization room in the TensorFlow framework, leading to a sub-optimal performance.

The sve-gemm optimization from \textit{blas-fp32} to \textit{sve-fp32} shows 1.3x performance improvement in the strong scaling scenario. Furthermore, the mixed-precision optimization accelerates the performance by 1.6x from double to MIX-fp32 and by 1.5x from MIX-fp32 to MIX-fp16. This proves that the mixed-precision method can significantly improve the performance of HPC applications while maintaining high accuracy. After adding the communication optimizations and the threadpool parallelization method in the \textit{comm\_nolb} experiments of the 1-2 atoms/core cases, the performance further improved by up to 22\%.

\subsection{Load Balance}

As shown in Fig. \ref{fig:compute_opt} \textit{comm\_lb} bars, the intra-node load balance improves the performance by a maximum of 18.5\%. We further analyze the pair time and atom number statistics of each MPI rank, shown in Table. \ref{tab:water_lb_distr}. We take the water system for example. As we mentioned above, the slowest MPI rank can slow down the whole simulation speed. Our load balance strategy successfully declines the max pair time by 16\% and 12\% in the 1-2 atoms/core cases, which is nearly the same growth as the overall performance. 

To measure the load balance, we use the standard deviation to mean ratio (SDMR), a metric used to measure the relationship between the volatility of the data and its average. SDMR is calculated through $SDMR=\sqrt{\frac{\sigma^2}{\mu } } \times 100$, where {$\sigma$} represents the variance, and {$\mu$} represents the average of the data. The higher the SDMR value, the greater the volatility. We can see the SDMR of pair time is cut in half, which reflects a great improvement in load balance.

The SDMR is 3x and 8x smaller in 1 atom/core and 2 atoms/core scenarios, respectively. However, the maximum count still exceeds our desired threshold, resulting in the busiest thread handling 2 atoms in the 1 atom/core case and 3 atoms in the 2 atoms/core case. Consequently, the SDMR still maintains a high degree even after load balance. That's why the copper system did not improve, as shown in Fig. \ref{fig:compute_opt} \textit{comm\_lb} bar with 1 atom/core. In practice, we can adjust the node scale according to the density condition of the physical system to achieve the highest simulation speed while saving computing resources. 

Finally, it's noteworthy that the difference in the pair phase time could also come from the ghost atom number of the local atom, system jitter, cache contention, and other uncontrollable factors. The analysis provided here primarily focuses on local atom numbers, which is a key factor in load balancing.

\begin{figure}[t]
  \centering
  \includegraphics[scale=0.7]{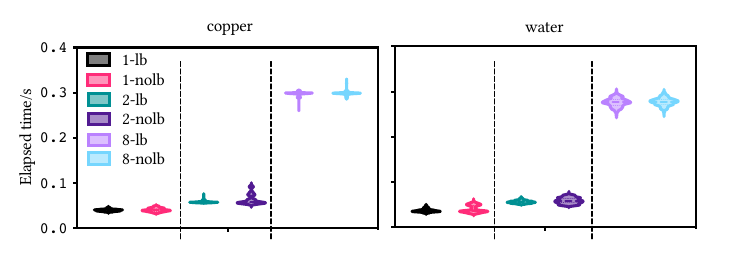}
  \caption{
  Pair time distribution. \textit{lb} and \textit{nolb} represent the load-balance version and the non-load-balance version, respectively. Additionally, the numbers 1, 2, and 8 stand for the atom number per core. Our optimized version significantly reduces the load imbalance between MPI ranks.
  }
  \label{fig:lb_pair_distribute}
\end{figure}

\begin{table}[t]
  \begin{center}
    \caption{Pair time and atom numbers across MPI ranks.}
    \label{tab:water_lb_distr}
    \begin{threeparttable}
    \begin{tabular}{lllllll}
      \hline
      \textbf{Case}                        & \textbf{lb}          & \textbf{type} & \textbf{Min}          & \textbf{Avg}          & \textbf{Max}          & \textbf{SDMR\%}         \\ \hline
      \multirow{4}{*}{\makecell{1 atom/core\\(12 atoms/rank)}}         & \multirow{2}{*}{no}  & pair          & $3.01$                & $4.11$                & $5.93$                & $3.89$  \\ 
                                           &                      & natom         & $7$                   & $11.625$              & $18$                  &  $79.93$  \\ \cline{2-7} 
                                           & \multirow{2}{*}{yes} & pair          & $3.05$                & $3.73$                & $4.98$                &  $1.85$  \\ 
                                           &                      & natom         & $10$                  & $11.625$              & $13$                  &  $24.32$  \\ \hline
      \multirow{4}{*}{\makecell{2 atoms/core\\(24 atoms/rank)}}         & \multirow{2}{*}{no}  & pair          & $4.77$                & $5.95$                & $7.60$                &  $2.77$  \\ 
                                           &                      & natom         & $16$                  & $23.25$               & $31$                  &  $90.81$  \\ \cline{2-7} 
                                           & \multirow{2}{*}{yes} & pair          & $5.07$                & $5.65$                & $6.69$                &  $1.27$  \\ 
                                           &                      & natom         & $22$                  & $23.25$               & $25$                  &  $11.10$  \\ \hline
      \multirow{4}{*}{\makecell{8 atoms/core\\(96 atoms/rank)}}         & \multirow{2}{*}{no}  & pair          & $24.98$               & $27.78$               & $30.05$               &  $1.65$  \\ 
                                           &                      & natom         & $87$                  & $93$                  & $97$                  &  $17.80$  \\ \cline{2-7} 
                                           & \multirow{2}{*}{yes} & pair          & $24.75$               & $27.64$               & $30.12$               &  $1.81$  \\ 
                                           &                      & natom         & $91$                  & $93$                  & $95$                  &  $9.32$  \\ \hline
        
    \end{tabular}
    \begin{tablenotes}    
        \footnotesize            
        \item[1] Pair is the pair phase time recorded by LAMMPS and the unit is 0.01s.
        \item[2] The column of $lb$ is whether open up the load-balance function.
        \item[3] SDMR stands for standard deviation to mean ratio.
    \end{tablenotes}         
  \end{threeparttable}
  \end{center}
\end{table}

\subsection{Strong scaling results}

We conduct strong scaling experiments on 540,000 copper atoms and 558,000 water molecule systems, scaling from 768 nodes to 12,000 nodes. The node topology settings are 8x12x8, 12x15x12, 16x18x16, 16x24x16, and 20x30x20 for 768, 2160, 4608, 6144, and 12,000 nodes. 

\begin{figure}[h]
  \centering
  \includegraphics[scale=0.65]{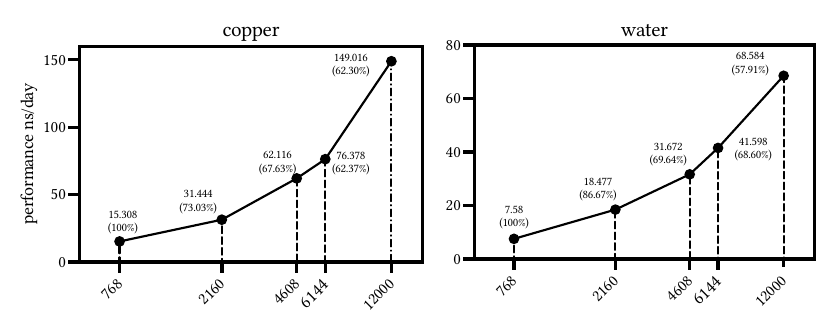}
  \caption{
  The strong scaling results of a 0.54M copper system and a 0.56M water system, scaling from 768 to 12,000 nodes. 
  Our optimized method achieves 149 nanoseconds per day for the copper system (31.7x speedup) and 68.5 nanoseconds per day for the water system (32.6x speedup) compared to the state-of-the-art work. 
  }
  \label{fig:strong_scaling}
\end{figure}

As shown in Fig. \ref{fig:strong_scaling}, our optimized code scales well and sustains high parallel efficiency, reaching 62.3\% and 57.9\% when scaling to 12,000 nodes. At the last data point, we achieve the simulation speed of 149 and 68.5 nanoseconds per day for the copper and water systems, respectively, which is the fastest simulation speed with ab initio accuracy as far as we know.

Besides, at the last data point, the average atoms per core stand at 0.93 and 0.968 for each system, respectively. According to the above analysis, the maximum atoms in a core may be 2 due to the load imbalance, leading to a slowdown in overall simulation time. If we continue to expand the node scale, the simulation speed can be upgraded further, which will make more cores idle. Thus, in practice, we should balance the simulation speed and economic efficiency according to the simulation scenario.

\section{Portability}

In this section, we discuss the portability of our optimizations. Our optimizations can be divided into 2 types, architecture-independent and architecture-dependent.

The computational optimizations for the DeePMD potential, including removing the TensorFlow framework, preprocessing the GEMM-NT operation, implementing sve-gemm, and employing mixed precision, are architecture-independent. These optimizations can be directly utilized by other machines.

However, other optimizations, including the node-based parallelization scheme and intra-node load balance, rely on Fugaku's unique features and cannot be directly applied to other machines. Nonetheless, we believe that with some engineering work, our optimizations could benefit other architectures as well. Details are as follows:

\begin{itemize}
\item \textbf{Node-based parallelization scheme}. The communication optimization essentially leverages intra-node communication to reduce inter-node communication and fully utilizes NICs (all the RDMA engines in Fugaku) with multi-threading to finish neighboring communication. Modern supercomputers typically feature multi-NUMA or multi-GPU architectures offering extremely high intra-node communication bandwidth and low latency, as well as multiple NICs for inter-node communication on each node. Examples include the many-core architecture machine of Sunway (NoC + 2x RDMA NICs) and the heterogeneous architecture machine of Frontier (Infinity Fabric + 4x Slingshots). Other machines can also reduce their communication time by leveraging GPU P2P or NoC to achieve low overhead intra-node gather and scatter operation, and extracting the full performance potential of the NICs with multi-threading inter-node communication. 
\item \textbf{Intra-node load-balance}. In the strong scaling scenario, the computation time difference between MPI ranks within a node can be hundreds of microseconds, significantly exceeding intra-node communication time. Therefore, by leveraging GPU P2P or NoC to evenly distribute atoms within a node, other machines can achieve performance improvement as well.
\end{itemize}

\section{Conclusion}

In this paper, we significantly enhance the strong scaling of the DeePMD-kit on the Fugaku supercomputer. We propose a node-based parallelization scheme that leverages the TofuD 6D torus/mesh network and its high-performance NoC, reducing communication overhead by 81\%. Furthermore, we introduce a series of computation optimizations, including TensorFlow simplification, mixed-precision, and high-performance sve-gemm, which collectively boost computation efficiency by a factor of 14.11x. Finally, we implement an intra-node load balance strategy, resulting in a 79.7\% reduction of atomic dispersion across all nodes. Testing results show that we achieve simulation speeds of 149 ns/day for copper and 68.5 ns/day for water on 12,000 nodes, successfully advancing the MD simulation with \textit{ab initio} accuracy to $>$100 ns/day for the first time.

\section*{Acknowledgment}

 This work is supported by the following funding: National Key Research and Development Program of China (2021YFB0300600), National Science Foundation of China (92270206, T2125013, 62372435, 62032023, 61972377,  61972380, T2293702), CAS Project for Young Scientists in Basic Research (YSBR-005) and the Strategic Priority Research Program of Chinese Academy of Sciences, Grant No. XDB0500102, China National Postdoctoral Program for Innovative Talents (BX20240383), and Huawei Technologies Co., Ltd.. This work used computational resources of the supercomputer Fugaku provided by RIKEN through the HPCI System Research Project (Project ID: hp240303). The numerical calculations in this study were partially carried out on the ORISE Supercomputer. We thank Prof. Xiaohui Duan (Shandong University) for helpful discussions.





\begin{thebibliography}{10}
\providecommand{\url}[1]{#1}
\csname url@samestyle\endcsname
\providecommand{\newblock}{\relax}
\providecommand{\bibinfo}[2]{#2}
\providecommand{\BIBentrySTDinterwordspacing}{\spaceskip=0pt\relax}
\providecommand{\BIBentryALTinterwordstretchfactor}{4}
\providecommand{\BIBentryALTinterwordspacing}{\spaceskip=\fontdimen2\font plus
\BIBentryALTinterwordstretchfactor\fontdimen3\font minus \fontdimen4\font\relax}
\providecommand{\BIBforeignlanguage}[2]{{%
\expandafter\ifx\csname l@#1\endcsname\relax
\typeout{** WARNING: IEEEtran.bst: No hyphenation pattern has been}%
\typeout{** loaded for the language `#1'. Using the pattern for}%
\typeout{** the default language instead.}%
\else
\language=\csname l@#1\endcsname
\fi
#2}}
\providecommand{\BIBdecl}{\relax}
\BIBdecl

\bibitem{ainsworth1984assessment}
R.~Ainsworth, ``The assessment of defects in structures of strain hardening material,'' \emph{Engineering Fracture Mechanics}, vol.~19, no.~4, pp. 633--642, 1984.

\bibitem{onuki2002phase}
A.~Onuki, \emph{Phase transition dynamics}.\hskip 1em plus 0.5em minus 0.4em\relax Cambridge University Press, 2002.

\bibitem{raty2005growth}
J.-Y. Raty, F.~Gygi, and G.~Galli, ``Growth of carbon nanotubes on metal nanoparticles: a microscopic mechanism from ab initio molecular dynamics simulations,'' \emph{Physical review letters}, vol.~95, no.~9, p. 096103, 2005.

\bibitem{callender2006advances}
R.~Callender and R.~B. Dyer, ``Advances in time-resolved approaches to characterize the dynamical nature of enzymatic catalysis,'' \emph{Chemical reviews}, vol. 106, no.~8, pp. 3031--3042, 2006.

\bibitem{anton}
D.~E. Shaw, R.~O. Dror, J.~K. Salmon, J.~Grossman, K.~M. Mackenzie, J.~A. Bank, C.~Young, M.~M. Deneroff, B.~Batson, K.~J. Bowers \emph{et~al.}, ``Millisecond-scale molecular dynamics simulations on anton,'' in \emph{Proceedings of the conference on high performance computing networking, storage and analysis}, 2009, pp. 1--11.

\bibitem{car1985unified}
R.~Car and M.~Parrinello, ``Unified approach for molecular dynamics and density-functional theory,'' \emph{Physical review letters}, vol.~55, no.~22, p. 2471, 1985.

\bibitem{jia2013analysis}
W.~Jia, Z.~Cao, L.~Wang, J.~Fu, X.~Chi, W.~Gao, and L.-W. Wang, ``The analysis of a plane wave pseudopotential density functional theory code on a gpu machine,'' \emph{Computer Physics Communications}, vol. 184, no.~1, pp. 9--18, 2013.

\bibitem{JIA2013102}
\BIBentryALTinterwordspacing
W.~Jia, J.~Fu, Z.~Cao, L.~Wang, X.~Chi, W.~Gao, and L.-W. Wang, ``Fast plane wave density functional theory molecular dynamics calculations on multi-gpu machines,'' \emph{Journal of Computational Physics}, vol. 251, pp. 102--115, 2013. [Online]. Available: \url{https://www.sciencedirect.com/science/article/pii/S002199911300329X}
\BIBentrySTDinterwordspacing

\bibitem{yanyujin2024jcst10millionatom}
\BIBentryALTinterwordspacing
Y.-J. Yan, H.-B. Li, T.~Zhao, L.-W. Wang, L.~Shi, T.~Liu, G.-M. Tan, W.-L. Jia, and N.-H. Sun, ``10-million atoms simulation of first-principle package ls3df,'' \emph{Journal of Computer Science and Technology}, vol.~39, no.~1, pp. 45--62, 2024. [Online]. Available: \url{https://jcst.ict.ac.cn/en/article/doi/10.1007/s11390-023-3011-6}
\BIBentrySTDinterwordspacing

\bibitem{jia_sc11_fast}
L.~Wang, Y.~Wu, W.~Jia, W.~Gao, X.~Chi, and L.-W. Wang, ``Large scale plane wave pseudopotential density functional theory calculations on gpu clusters,'' in \emph{SC '11: Proceedings of 2011 International Conference for High Performance Computing, Networking, Storage and Analysis}, 2011, pp. 1--10.

\bibitem{behler2007generalized}
J.~Behler and M.~Parrinello, ``Generalized neural-network representation of high-dimensional potential-energy surfaces,'' \emph{Physical review letters}, vol.~98, no.~14, p. 146401, 2007.

\bibitem{THOMPSON2015316}
\BIBentryALTinterwordspacing
A.~Thompson, L.~Swiler, C.~Trott, S.~Foiles, and G.~Tucker, ``Spectral neighbor analysis method for automated generation of quantum-accurate interatomic potentials,'' \emph{Journal of Computational Physics}, vol. 285, pp. 316--330, 2015. [Online]. Available: \url{https://www.sciencedirect.com/science/article/pii/S0021999114008353}
\BIBentrySTDinterwordspacing

\bibitem{lee2019simple}
K.~Lee, D.~Yoo, W.~Jeong, and S.~Han, ``Simple-nn: An efficient package for training and executing neural-network interatomic potentials,'' \emph{Computer Physics Communications}, vol. 242, pp. 95--103, 2019.

\bibitem{Behler_2014}
\BIBentryALTinterwordspacing
J.~Behler, ``Representing potential energy surfaces by high-dimensional neural network potentials,'' \emph{Journal of Physics: Condensed Matter}, vol.~26, no.~18, p. 183001, apr 2014. [Online]. Available: \url{https://doi.org/10.1088/0953-8984/26/18/183001}
\BIBentrySTDinterwordspacing

\bibitem{PhysRevLett.98.146401}
\BIBentryALTinterwordspacing
J.~Behler and M.~Parrinello, ``Generalized neural-network representation of high-dimensional potential-energy surfaces,'' \emph{Phys. Rev. Lett.}, vol.~98, p. 146401, Apr 2007. [Online]. Available: \url{https://link.aps.org/doi/10.1103/PhysRevLett.98.146401}
\BIBentrySTDinterwordspacing

\bibitem{behler2017first}
J.~Behler, ``First principles neural network potentials for reactive simulations of large molecular and condensed systems,'' \emph{Angewandte Chemie International Edition}, vol.~56, no.~42, pp. 12\,828--12\,840, 2017.

\bibitem{doi:10.1021/acs.jpclett.7b01072}
\BIBentryALTinterwordspacing
K.~Yao, J.~E. Herr, S.~N. Brown, and J.~Parkhill, ``Intrinsic bond energies from a bonds-in-molecules neural network,'' \emph{The Journal of Physical Chemistry Letters}, vol.~8, no.~12, pp. 2689--2694, 2017, pMID: 28573865. [Online]. Available: \url{https://doi.org/10.1021/acs.jpclett.7b01072}
\BIBentrySTDinterwordspacing

\bibitem{smith2017ani}
J.~S. Smith, O.~Isayev, and A.~E. Roitberg, ``Ani-1: an extensible neural network potential with dft accuracy at force field computational cost,'' \emph{Chemical science}, vol.~8, no.~4, pp. 3192--3203, 2017.

\bibitem{DESAI2022108156}
\BIBentryALTinterwordspacing
S.~Desai, S.~T. Reeve, and J.~F. Belak, ``Implementing a neural network interatomic model with performance portability for emerging exascale architectures,'' \emph{Computer Physics Communications}, vol. 270, p. 108156, 2022. [Online]. Available: \url{https://www.sciencedirect.com/science/article/pii/S001046552100268X}
\BIBentrySTDinterwordspacing

\bibitem{https://doi.org/10.1002/cjoc.202100456}
\BIBentryALTinterwordspacing
Y.-P. Huang, Y.~Xia, L.~Yang, J.~Wei, Y.~I. Yang, and Y.~Q. Gao, ``Sponge: A gpu-accelerated molecular dynamics package with enhanced sampling and ai-driven algorithms,'' \emph{Chinese Journal of Chemistry}, vol.~40, no.~1, pp. 160--168, 2022. [Online]. Available: \url{https://onlinelibrary.wiley.com/doi/abs/10.1002/cjoc.202100456}
\BIBentrySTDinterwordspacing

\bibitem{wang2018deepmd}
H.~Wang, L.~Zhang, J.~Han, and E.~Weinan, ``Deepmd-kit: A deep learning package for many-body potential energy representation and molecular dynamics,'' \emph{Computer Physics Communications}, vol. 228, pp. 178--184, 2018.

\bibitem{schutt2017schnet}
K.~Sch{\"u}tt, P.-J. Kindermans, H.~E. Sauceda~Felix, S.~Chmiela, A.~Tkatchenko, and K.-R. M{\"u}ller, ``Schnet: A continuous-filter convolutional neural network for modeling quantum interactions,'' \emph{Advances in neural information processing systems}, vol.~30, 2017.

\bibitem{PhysRevB.99.014104}
\BIBentryALTinterwordspacing
R.~Drautz, ``Atomic cluster expansion for accurate and transferable interatomic potentials,'' \emph{Phys. Rev. B}, vol.~99, p. 014104, Jan 2019. [Online]. Available: \url{https://link.aps.org/doi/10.1103/PhysRevB.99.014104}
\BIBentrySTDinterwordspacing

\bibitem{Lysogorskiy2021}
\BIBentryALTinterwordspacing
Y.~Lysogorskiy, C.~v.~d. Oord, A.~Bochkarev, S.~Menon, M.~Rinaldi, T.~Hammerschmidt, M.~Mrovec, A.~Thompson, G.~Cs{\'a}nyi, C.~Ortner, and R.~Drautz, ``Performant implementation of the atomic cluster expansion (pace) and application to copper and silicon,'' \emph{npj Computational Materials}, vol.~7, no.~1, p.~97, Jun 2021. [Online]. Available: \url{https://doi.org/10.1038/s41524-021-00559-9}
\BIBentrySTDinterwordspacing

\bibitem{Batzner2022}
\BIBentryALTinterwordspacing
S.~Batzner, A.~Musaelian, L.~Sun, M.~Geiger, J.~P. Mailoa, M.~Kornbluth, N.~Molinari, T.~E. Smidt, and B.~Kozinsky, ``E(3)-equivariant graph neural networks for data-efficient and accurate interatomic potentials,'' \emph{Nature Communications}, vol.~13, no.~1, p. 2453, May 2022. [Online]. Available: \url{https://doi.org/10.1038/s41467-022-29939-5}
\BIBentrySTDinterwordspacing

\bibitem{gasteiger_dimenet_2020}
J.~Gasteiger, J.~Gro{\ss}, and S.~G{\"u}nnemann, ``Directional message passing for molecular graphs,'' in \emph{International Conference on Learning Representations (ICLR)}, 2020.

\bibitem{gasteiger_dimenetpp_2020}
J.~Gasteiger, S.~Giri, J.~T. Margraf, and S.~G{\"u}nnemann, ``Fast and uncertainty-aware directional message passing for non-equilibrium molecules,'' in \emph{Machine Learning for Molecules Workshop, NeurIPS}, 2020.

\bibitem{Unke2021}
\BIBentryALTinterwordspacing
O.~T. Unke, S.~Chmiela, M.~Gastegger, K.~T. Sch{\"u}tt, H.~E. Sauceda, and K.-R. M{\"u}ller, ``Spookynet: Learning force fields with electronic degrees of freedom and nonlocal effects,'' \emph{Nature Communications}, vol.~12, no.~1, p. 7273, Dec 2021. [Online]. Available: \url{https://doi.org/10.1038/s41467-021-27504-0}
\BIBentrySTDinterwordspacing

\bibitem{musaelian2023learning}
A.~Musaelian, S.~Batzner, A.~Johansson, L.~Sun, C.~J. Owen, M.~Kornbluth, and B.~Kozinsky, ``Learning local equivariant representations for large-scale atomistic dynamics,'' \emph{Nature Communications}, vol.~14, no.~1, p. 579, 2023.

\bibitem{zhang2023dpa}
D.~Zhang, X.~Liu, X.~Zhang, C.~Zhang, C.~Cai, H.~Bi, Y.~Du, X.~Qin, J.~Huang, B.~Li \emph{et~al.}, ``Dpa-2: Towards a universal large atomic model for molecular and material simulation,'' \emph{arXiv preprint arXiv:2312.15492}, 2023.

\bibitem{musaelian2023scaling}
A.~Musaelian, A.~Johansson, S.~Batzner, and B.~Kozinsky, ``Scaling the leading accuracy of deep equivariant models to biomolecular simulations of realistic size,'' \emph{arXiv preprint arXiv:2304.10061}, 2023.

\bibitem{nguyen2021billion}
K.~Nguyen-Cong, J.~T. Willman, S.~G. Moore, A.~B. Belonoshko, R.~Gayatri, E.~Weinberg, M.~A. Wood, A.~P. Thompson, and I.~I. Oleynik, ``Billion atom molecular dynamics simulations of carbon at extreme conditions and experimental time and length scales,'' in \emph{Proceedings of the International Conference for High Performance Computing, Networking, Storage and Analysis}, 2021, pp. 1--12.

\bibitem{guo2022extending}
Z.~Guo, D.~Lu, Y.~Yan, S.~Hu, R.~Liu, G.~Tan, N.~Sun, W.~Jiang, L.~Liu, Y.~Chen \emph{et~al.}, ``Extending the limit of molecular dynamics with ab initio accuracy to 10 billion atoms,'' in \emph{Proceedings of the 27th ACM SIGPLAN Symposium on Principles and Practice of Parallel Programming}, 2022, pp. 205--218.

\bibitem{jia2020pushing}
W.~Jia, H.~Wang, M.~Chen, D.~Lu, L.~Lin, R.~Car, E.~Weinan, and L.~Zhang, ``Pushing the limit of molecular dynamics with ab initio accuracy to 100 million atoms with machine learning,'' in \emph{SC20: International conference for high performance computing, networking, storage and analysis}.\hskip 1em plus 0.5em minus 0.4em\relax IEEE, 2020, pp. 1--14.

\bibitem{zeng2023deepmd}
J.~Zeng, D.~Zhang, D.~Lu, P.~Mo, Z.~Li, Y.~Chen, M.~Rynik, L.~Huang, Z.~Li, S.~Shi \emph{et~al.}, ``Deepmd-kit v2: A software package for deep potential models,'' \emph{The Journal of Chemical Physics}, vol. 159, no.~5, 2023.

\bibitem{galib2020elucidating}
M.~Galib and D.~T. Limmer, ``Elucidating the mechanism of reactive uptake of n $ \_2 $ o $ \_5 $ in aqueous aerosol,'' \emph{arXiv preprint arXiv:2005.10134}, 2020.

\bibitem{du2023efficient}
S.~Du, X.~You, H.~Yang, J.~Shang, Z.~Xiao, Z.~Wu, Z.~Luan, and D.~Qian, ``Efficient deep molecular dynamic model training on heterogeneous system,'' in \emph{2023 IEEE 29th International Conference on Parallel and Distributed Systems (ICPADS)}.\hskip 1em plus 0.5em minus 0.4em\relax IEEE, 2023, pp. 1869--1876.

\bibitem{singraber2019parallel}
A.~Singraber, T.~Morawietz, J.~Behler, and C.~Dellago, ``Parallel multistream training of high-dimensional neural network potentials,'' \emph{Journal of chemical theory and computation}, vol.~15, no.~5, pp. 3075--3092, 2019.

\bibitem{duan2020cell}
X.~Duan, P.~Gao, M.~Zhang, T.~Zhang, H.~Meng, Y.~Li, B.~Schmidt, H.~Fu, L.~Gan, W.~Xue \emph{et~al.}, ``Cell-list based molecular dynamics on many-core processors: a case study on sunway taihulight supercomputer,'' in \emph{SC20: International Conference for High Performance Computing, Networking, Storage and Analysis}.\hskip 1em plus 0.5em minus 0.4em\relax IEEE, 2020, pp. 1--12.

\bibitem{li2023enhance}
J.~Li, T.~Zhao, Z.~Guo, S.~Shi, L.~Liu, G.~Tan, W.~Jia, G.~Yuan, and Z.~Wang, ``Enhance the strong scaling of lammps on fugaku,'' in \emph{Proceedings of the International Conference for High Performance Computing, Networking, Storage and Analysis}, 2023, pp. 1--13.

\bibitem{top500_fugaku}
{TOP500.org}, ``Supercomputer fugaku - supercomputer fugaku, a64fx 48c 2.2ghz, tofu interconnect d,'' \url{https://www.top500.org/system/179807/}, 2023.

\bibitem{lu2022dp}
D.~Lu, W.~Jiang, Y.~Chen, L.~Zhang, W.~Jia, H.~Wang, and M.~Chen, ``Dp compress: A model compression scheme for generating efficient deep potential models,'' \emph{Journal of chemical theory and computation}, vol.~18, no.~9, pp. 5559--5567, 2022.

\bibitem{ko2019_water}
H.-Y. Ko, L.~Zhang, B.~Santra, H.~Wang, W.~E, R.~A. DiStasio~Jr, and R.~Car, ``Isotope effects in liquid water via deep potential molecular dynamics,'' \emph{Molecular Physics}, vol. 117, no.~22, pp. 3269--3281, 2019.

\bibitem{distasio2014_water}
R.~A. DiStasio, B.~Santra, Z.~Li, X.~Wu, and R.~Car, ``The individual and collective effects of exact exchange and dispersion interactions on the ab initio structure of liquid water,'' \emph{The Journal of chemical physics}, vol. 141, no.~8, 2014.

\bibitem{chen2017_water}
M.~Chen, H.-Y. Ko, R.~C. Remsing, M.~F. Calegari~Andrade, B.~Santra, Z.~Sun, A.~Selloni, R.~Car, M.~L. Klein, J.~P. Perdew \emph{et~al.}, ``Ab initio theory and modeling of water,'' \emph{Proceedings of the National Academy of Sciences}, vol. 114, no.~41, pp. 10\,846--10\,851, 2017.

\bibitem{zhang2018deep}
L.~Zhang, J.~Han, H.~Wang, R.~Car, and E.~Weinan, ``Deep potential molecular dynamics: a scalable model with the accuracy of quantum mechanics,'' \emph{Physical review letters}, vol. 120, no.~14, p. 143001, 2018.

\bibitem{3-stage_pattern_desmond}
K.~J. Bowers, E.~Chow, H.~Xu, R.~O. Dror, M.~P. Eastwood, B.~A. Gregersen, J.~L. Klepeis, I.~Kolossvary, M.~A. Moraes, F.~D. Sacerdoti \emph{et~al.}, ``Scalable algorithms for molecular dynamics simulations on commodity clusters,'' in \emph{Proceedings of the 2006 ACM/IEEE Conference on Supercomputing}, 2006, pp. 84--es.

\end{thebibliography}


\end{document}